\documentclass[aps,twocolumn]{revtex4}
\usepackage{amssymb,amsmath,epsfig}
\def\av#1{\left\langle#1\right\rangle}
\begin{document}

\title{Search in Complex Networks:
a New Method of Naming}


\author{Shai Carmi$^1$, Reuven Cohen$^2$ and Danny Dolev$^3$}

\affiliation{$^1$Minerva Center and Dept. of Physics,
Bar-Ilan University, Ramat-Gan, Israel\\
$^2$Electrical and Computer Engineering Department, Boston University,
Boston, MA, USA\\
$^3$School of Engineering and Computer Science, Hebrew University,
Jerusalem, Israel}

\begin{abstract}
We suggest a method for routing when the source does
not posses full information about the shortest path to the
destination. The method is particularly useful for scale-free
networks, and exploits its unique characteristics. By assigning
new (short) names to nodes (aka labelling) we are able to reduce
significantly the memory requirement at the routers, yet we
succeed in routing with high probability through paths very close
in distance to the shortest ones.
\end{abstract}


\maketitle

In recent years it has been shown that many real world networks,
such as technological, social and biological networks, and in particular,
the Internet, are scale free, i.e. have a
power law degree distribution \cite{BA,FFF,vesp_book,dor_book}. 
The probability of a site to have
degree $k$, $P(k)\sim k^{-\gamma}$, where in the Internet, it is
assumed that $\gamma\approx 2.1 - 2.5$. One of the most important
tasks in networking is routing. Efficient routing is necessary in
order to provide efficient transportation and utilization of the
network resources. In the context of communication networks,
routing is an important task in packet switched networks as well
as in overlay networks (such as Peer-to-Peer).
In this paper we present a method for
searching for nodes and routing where no knowledge of the location
of the destination node is given. Such methods are usually known
as ``compact routing'' schemes.

In order to obtain good results several variables should be considered:
\begin{itemize}
\item
The stretch is defined as the ratio of the actual routing path to the
shortest path
between two given nodes. The smallest the ratio the more efficient the
communication in the network.
\item
The table size is the number of entries kept in the storage of each
node. The smaller the table the more efficient the scheme in
terms of memory requirements.
\item
The label size is the number of bits presenting the name (or
address) of each node. The smallest possible label size needed to
distinguish between sites with a unique id is logarithmic. Most efficient
routing schemes use larger labels in order to present more
information about the node.
\end{itemize}

In many cases it is desirable to design an approximate routing schemes
that require
considerably smaller tables, in the cost of allowing for higher stretch
(shortest path routing not guaranteed), and larger labels.

Partial knowledge search in a small-world lattice based network and
power-law networks
was investigated in \cite{kleinberg,huberman,sneppen}.
The first work on generalized routing with a tradeoff of table size
vs. label size and stretch was given by Peleg and
Upfal~\cite{peleg}. This scheme has later been extended by Thorup
and Zwick~\cite{zwick} and by Cowen~\cite{cowen}. All those schemes
require a rather large table (of order $N^{1/2}$ to ensure an upper
bound of 3 for the stretch,
or, in general, $O(N^\frac{2}{s+1})$ for an odd stretch $s$).
A numerical study of the actual stretch for scale free networks is
presented in~\cite{fall}, showing that the actual performance of the
above routing schemes, in terms of the average stretch, is much
better than the worst case guarantee.

In this paper we discuss a class of routing schemes with a
parameter $H$ ($1\leq H\leq N$), which is proportional to the
memory requirement at the nodes.
We give arguments showing that the ratio of the average routing distance to the average
shortest path is below 2 with high probability, mo matter what $H$ is.
For scale-free networks the stretch
is usually much lower, and we show analytically and numerically
that \emph{even} for very small values of $H$, $H =
O(\log^{\nu}N)$ for $\nu \geq 0$, the actual stretch is very close
to 1. Thus, a routing scheme that requires substantially small
tables and poly-logarithmic labels (see below) may lead to a very
efficient routing. When comparing properly, our scheme is more
efficient than previous ones; moreover, our scheme is simpler and
more intuitive (e.g. do not involve randomization), and the
trade-off between performance and memory requirements is
controllable.

The random network model we use here is the Configuration Model
of~\cite{bol}. The networks in this model are created by the
following process: given a network with $N$ nodes,
and a degree sequence $k_{i, 1 \leq i \leq N}$,
create a list containing $k_i$ copies of each
node $i$, and choose a random matching on this list to create
the edges of the network. We ignore self loops and multiple edges,
which are statistically insignificant \cite{newman}.

The main degree sequence we will discuss is of scale free
networks: $P(k)\sim k^{-\gamma}$, (with $k\geq k_{min}$). This
degree sequence has been shown to exist naturally in many
networks~\cite{BA}, in particular, the Internet~\cite{FFF} and P2P
networks~\cite{P2P} as discussed above. Another degree sequence
which we will use for comparison is the one of the
Erd{\"o}s-R{\'e}nyi (ER) random network model, $P(k)=
\frac{e^{-\gamma}\gamma^k}{k!}$.

The proposed routing scheme consists of two stages: the preprocessing and
the actual routing.
\begin{description}
\item {\bf{Preprocessing}}
The $H$ highest degree nodes are designated as the ``hubs''. (Ties
in the degree are broken arbitrarily). For each site $i$ the
closest hub $h_i$ is searched (ties are broken by degree).
Designate the shortest path from site $i$ to its hub $h_i$ \mbox{by --}\\
$v_{i0},v_{i1},v_{i2},\ldots,v_{i,{n_i}},$ where $i=v_{i,{0}}$ and
$v_{i,{n_i}}=h_i$. The label for site $i$ will be $L_i=\langle
i,v_{i,1},v_{i,2},\ldots,v_{i,{n_i}-1},h_i\rangle$. The routing
table for each node in the network contains the link leading to the
shortest path for each of the hubs, as well as a list of all of its
immediate neighbors.
\item {\bf{Actual Routing}}
Assume a packet is sent from some initial node towards the
destination node $t$. As the packet reaches some intermediate node
$x$, it is handled by the following algorithm:
\begin{enumerate}
\item
If $x=t$ then stop.
\item
If $t$ is a neighbor of $x$, then send the packet directly to $t$.
\item
Otherwise, if $x \in L_t$, i.e. $x=v_{t,j}$ for some $j,$ then move
the packet to $v_{t,j-1}$.
\item
Otherwise, search for $h_t$ in the table and send the packet through
the appropriate link.
\end{enumerate}
\end{description}

Let us first show that our method is efficient by means of average
running time.

\emph{Preprocessing}: Choosing the $H$ hubs and sorting them can
be done in $O(N + H\log H)$ time \cite{cormen}. Next, from each
hub we need only to start a Breadth First Search, keeping for each
node $x$ that is reached its distance to the root and its
predecessor (storing those in $x$'s routing table). Next for each
node we decide which is the closest hub, find the path to that
hub, and store it as its new label. All of this can be done in
$O(MH)$ time, where $M$ is the total number of edges (which is of
the order of $N$ in practical cases). Note that this running-time
is better than in previously suggested schemes \cite{zwick}.

\emph{Routing decision}: In each decision we need to search either
the label or the routing table. In practical cases the label size
is extremely small and can be considered constant; the routing
table can be implemented as a hash table to provide average
constant access time \cite{cormen}. Therefore we conclude that an
average routing decision can be done in constant time.

We now look at the average distance travelled by a packet relative
to the average shortest path in the network. The average is taken
over all pairs and all configurations of the network in the network
model presented above.

We use the following lemma.
Let $a_1$ and $a_2$ be nodes with respective degrees \mbox{$k_{a_1}\geq
k_{a_2}$}, and $b$ be any other random node. Denote by $d(a,b)$ the
length of the shortest path between nodes $a$ and $b$, then we
claim that
\begin{equation}
\label{Peq}
P(d(a_1,b)\leq l) \geq P(d(a_2,b)\leq l)
\end{equation}
for all $l$.

To see that, we consider only cases in which the paths
$a_1\rightarrow b$ and $a_2\rightarrow b$ exist (otherwise the
distance is not defined). Now fix the connections in the sub-network
formed by deleting $a_1$ and $a_2$ from the original network,
and consider the links between this sub-network and $\{a_1,a_2\}$.
Assume that $p$ of the links lead to paths of length $l,$
which is the length of the shortest path to $b.$

If the network is with high probability fully connected (as
in random networks in which all degrees are at least $3$ \cite{saberi},
and the case of the Internet), then the
ratio of matchings for which $d(a_1,b)=l$ and $d(a_2,b)>l$ to those
where $d(a_1,b)>l$ and $d(a_2,b)=l$ is $\binom{k_{a_1}}{p} /
\binom{k_{a_2}}{p}$, and therefore the distance is a non-increasing
function of the degree.

In cases where the network is not fully connected, we must
condition the relevant matchings on the demand that both $a_1$ and
$a_2$ are connected to $b$. It can be shown that also in these
cases Eq. (\ref{Peq}) is valid. Therefore we conclude that,
$\av{d(a,b)}$, for some random node $b$, is a non-increasing
function of $k_a$ --
\begin{equation}
\label{degeq}
\forall a_1,a_2,b ~-~  k_{a_1} \leq k_{a_2} \Rightarrow \av{d(a_2,b)}
\leq \av{d(a_1,b)}
\end{equation}

Next we use the notation $d(a,b)$ for the length of the shortest
path between nodes $a$ and $b$, and $r(a,b)$ for the distance
travelled by a packet sent from $a$ to $b$ using the above
algorithm (notice that $r(a,b)$ need not be symmetric, as opposed
to $d(a,b)$ ). We argue, that in the proposed routing scheme,
the expected average stretch $S
\equiv \frac{\av{r(a,b)}}{\av{d(a,b)}}\leq 2$.

Denote the source node as $s$, the destination as $t$, the hub of $t$ as $h_t$,
and the lengths of the direct paths between them $d(s,t), d(s,h_t), d(t,h_t)$.
By the construction of the scheme:
\begin{eqnarray}
\label{Seq}
S &=& \frac{\av{r(s,t)}}{\av{d(s,t)}} \leq
\frac{\av{{d(s,h_t)+d(h_t,t)}}}{\av{d(s,t)}} \nonumber\\ &=&
\frac{\av{d(s,h_t)}}{\av{d(s,t)}} + \frac{\av{d(h_t,t)}}{\av{d(s,t)}}.
\end{eqnarray}

Consider first the case that the hub $h_t$ is just a random node, call it $r$.
Becasue of symmetry, there no reason why any of the distances
$d(s,t)$ , $d(s,r)$ , $d(r,t)$ would be larger than the other,
therefore on average the total routing distance $d(s,r)+d(r,t)$
is just twice the shortest distance $d(s,t)$,
or the average stretch is 2.

This is true for any random node being a hub,
but we are choosing the hubs as nodes with high degree.
Since eq. (\ref{degeq}) states that the average distance between a random node
and a hub is smaller than the distance between two random nodes,
we expect the average distances to and from the hub to be small,
i.e. we expect $d(s,h_t) \leq d(s,t)$ and $d(h_t,t) < d(s,t)$,
thus we expect that the average stertch $S \leq 2$.

(The cases in which $k_s,k_t > k_{h_t}$ are treated easily --
Since $h_t$ is the hub of $t$, then even if $s$ is a hub then
by the definition of the scheme $h_t$ is closer to $t$ than $s$, and $d(h_t,t) \leq
d(s,t)$;
if $t$ is a hub the routing is shortest path by construction.
Thus we can assume that $s$ and $t$ are not hubs and $k_s,k_t \leq k_{h_t}$).

Note that direct application of eq. (\ref{degeq}) is not possible since in the
derivation
we assumed the three nodes $\{a_1,a_2,b\}$ are fixed, while in our case rewiring
might cause $h_t$ not to be the hub closest to $t$ anymore.
Nevertheless, there is no reason to assume the inequalities will
be invalid for the reduced configuration space where we force $h_t$ to be
the hub closest to $t$.
Paradoxically, if there is only one hub, then the three nodes are fixed
and we can apply eq. (\ref{degeq}) directly, to prove $S<=2$. It is however obvious,
and confirmed by simulations, that increasing $H$ would decrease the stretch.

Other properties of the proposed scheme are:
\begin{enumerate}
\item
The label size (in bits) for the proposed scheme is at most $(D+1)\log N$,
where $D$ is the diameter of the network.
\item
The table size at every node contains $H + k$ entries, where $k$ is
the degree.
\item
The contents of the packet need not to be changed through the
routing process.
\item
The scheme is a shortest-path routing for a tree.
\end{enumerate}

To explain $1$, recall that the label contains the shortest path
to the closest hub. The distance is at most $D$ (and add one for
the site itself), and each node requires at most $\log N$ bits to
identify. Thus, property 1 follows. The second and third parts
follow from the definition of the scheme. The fourth follows since
in a tree there is only one path between any two nodes, so either
the hub is on the path, or the destination is on the path to the
hub, or there exists some node in the path to the hub which is
also on the path to the destination. (In a different way, if there
was a shortest path different from the path $source \rightarrow
hub \rightarrow destination$, then a loop would be constructed,
contradicting the network being a tree).

For scale-free networks we can show some better bounds on the
label size and the stretch. It has been shown~\cite{CH03,CL03}
that with high probability the average distance between nodes is
$O(\log\log{N})$ and the diameter is $O(\log{N})$ (for $k_{min}
\geq 2$ the diameter is also expected to be $O(\log\log{N})$).
Therefore, it can be concluded that the maximum label size is of
order $O(\log^2 N)$ and the average label size is
$O(\log{N}\log{\log{N}})$. For scale free networks with $\gamma<3$,
tighter bound for the stretch can be obtained. The radius of the
core (the location of all high degree nodes) is of order $\log\log
N$, and almost all the mass is concentrated outside the core (see,
e.g., \cite{CH03,hof}). Now, looking at a ball around a random
site with a radius a little smaller than the radius of the
network, it is expected that the ball will not include the largest
hub (since most sites are outside the core). Since the size of the
largest hub is of order $O(N^{1/(\gamma-1)})\gg
N^{1/2}$~\cite{CEBH00} for $\gamma<3,$ it is expected that the
ball has less than $N^{1/2}$ outgoing links (since any 2 balls
with more than $N^{1/2}$ are connected with high probability). Any
2 such balls are not expected to be connected between them, since
the product of their ``degree'' (number of outgoing links) is less
than $N$, so the distance between any two random sites is expected
to be almost twice the radius (for a rigorous proof of this
see~\cite{hof}). Thus the path through the hubs is almost optimal
with high probability, and the stretch between 2 randomly selected
sites is expected to approach 1 for large $N$.

One other nice property of the proposed scheme is that the
labelling and table construction can be achieved using a
distributed rather than a centralized algorithm, and in an
efficient manner (the number of messages transfers needed is
almost linear in the size of network times $H$. Details are to be
published elsewhere). Cases of a node or link failures can be
bypassed in a standard way, without affecting the other nodes of
the network. Having all the above properties in mind, our scheme
can be considered seriously for applications in real-world
systems, in which not always there is a central management of the
network that has the knowledge of the topology of the entire
network.

\begin{figure}
\epsfxsize=2.2in \epsfbox{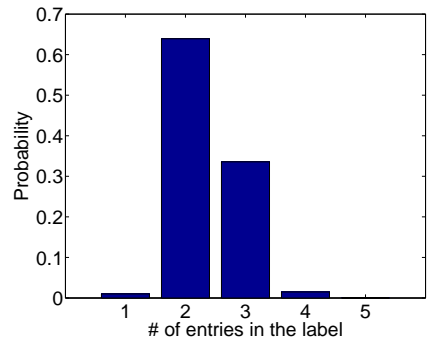} \caption{Label size
distribution for scale-free networks. Typical label is extremely
short, what makes our scheme efficient also in terms of bandwidth
utilization. \label{fig:label_dist}}
\end{figure}

To demonstrate the efficiency of the scheme, we present computer
simulation results. For all networks, we use the parameters
$N=10000$, $\gamma =2.3$, and average over many realizations. (The
stretch of a network is calculated as an average over the stretch
of all pairs, as in \cite{fall}). To begin with, we verify that
the labels are indeed small (Fig. \ref{fig:label_dist}).

\begin{figure}
\epsfxsize=2.2in
\epsfbox{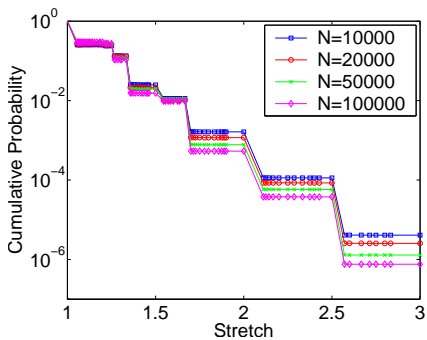}
\caption{Stretch distribution for a scale-free network (\mbox{$H \sim \log(N)$}).
The (inverse) cumulative probability distribution is shown,
i.e. for a given stretch value, we see the
probability to have a larger stretch.
In the case of $N=10000$, $75\%$ of the paths are the shortest ones.\label{fig:str_dist}}
\end{figure}

Next we have tested the scheme with the most recent representation
of the Internet at the AS level \cite{dimes}; the average stretch
factor turned out to be as low as $1.067$, with $79\%$ of paths
shortest (As opposed to $1.09$ and $71\%$ in \cite{fall}). In Fig.
\ref{fig:str_dist} we show the cumulative distribution of stretch
values for routing between all pairs in a random realization of
the configuration model (with power-law degree distribution), for
different system sizes. It can be seen that not only that most of
the routes are along the shortest path, but the number of
exceptionally high stretches becomes more and more rare as the
system grows.

\begin{figure}
\epsfxsize=2.2in \epsfbox{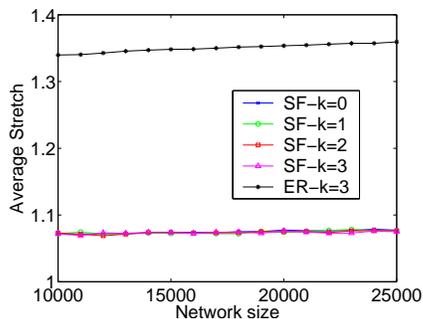} \caption{Average stretch vs.
network size, for scale-free networks with different number of
hubs, $H \sim \log^{\nu}N$, $\nu=0,1,2,3$ and ER network
($\av{k}=7$) with $\nu=3$. In all simulations H was scaled such
that $H(N=10000)=100$. It can be seen that the performance of the
scheme is much better for the scale-free network, with virtually
no dependence in the network size and the number of hubs.}
\label{fig:avg_str_N}
\end{figure}

Fig. \ref{fig:avg_str_N} shows the average stretch value as a
function of the network size, compared for a few values of $\nu$
\mbox{(in $H \sim \log^{\nu}N$)} in power-law networks, and for $H
\sim \log^3 N$ for ER networks. It can be seen that the average
stretch in the scale-free networks is significantly better than in
the ER case and is virtually independent of the network size. One
can also see that the stretch depends only weakly on the number of
hubs; therefore, to achieve an efficient routing, one need not use
too many hubs.

\begin{figure}
\epsfxsize=2.8in \epsfbox{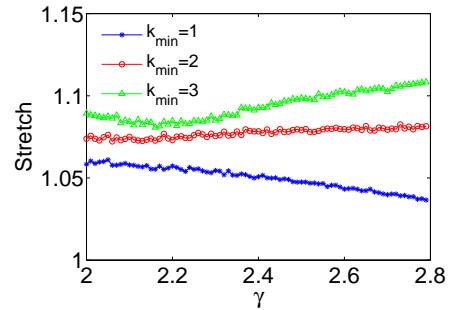} \caption{Average stretch
vs. $\gamma$ for a power-law network with $k_{min} = 1,2,3$.
\label{fig:avg_str_gamma}}
\end{figure}

In Fig. \ref{fig:avg_str_gamma} we study the variation in the
stretch when the parameters of the power-law degree distribution
are changed. We compute the stretch for $k_{min} = 1,2,3$ and for
various values of $\gamma$. The behaviour of the stretch can be
explained, as when we move to higher values of $\gamma$, the
network becomes more sparse and tree-like. On the one hand recall
that the scheme is optimal for tree structure, on the other hand
when $\gamma$ increases we have less and less ``real hubs``, the
network becomes similar to an ER network, on which the scheme
performs worse, as shown above. For $k_{min}=1$ the tree structure
effect is much stronger, for $k_{min}=3$ many loops remain thus
the effect of losing the hubs is stronger, for $k_{min}=2$ neither
of the effects is more significant.

In summary, we have presented an efficient method for routing or
searching in an environment where full knowledge of the network
topology is not available. Our scheme changes the names of the
nodes to more meaningful names, that contain the path to the
closest hub, where the hubs are chosen as nodes with highest
degree. We have shown that this simple and intuitive method can be
extremely useful in scale-free networks, such as the Internet.
Using computer simulations, we have explored the performance of
our scheme with variations in the network and scheme parameters.

We thank Shlomo Havlin and Ittai Abraham
for fruitful discussions. The research was
supported by the Israel Internet Association, the Israel Science
Foundation and the European research NEST/PATHFINDER project
DYSONET 012911.




\end{document}